\documentclass{mn2e}

\usepackage{graphicx}

\newcommand{\beq}{\begin{equation}}
\newcommand{\eeq}{\end{equation}}
\newcommand{\lab}{\label}
\newcommand{\ba}{\begin{array}}
\newcommand{\ea}{\end{array}}

\newcommand{\bfxi}{\mbox{\boldmath $\xi$}}

\newcommand{\bfal}{\mbox{\boldmath $\alpha$}}
\newcommand{\bal}{\mbox{\boldmath $\alpha$}}
\newcommand{\bx}{{\bf x}}
\newcommand{\by}{{\bf y}}

\begin{document}

\title{Gravitational lensing by stars with angular momentum}

\author[M. Sereno]
{M. Sereno,\thanks{E-mail: Mauro.Sereno@na.infn.it.}
\\Dipartimento di Scienze Fisiche, Universit\`{a} degli Studi di Napoli
``Federico II", \\and Istituto Nazionale di Fisica Nucleare, Sec.
Napoli,\\Compl. Univ. Monte S. Angelo, via Cinthia, 45--80126, Napoli,
Italia.}
\maketitle

\begin{abstract}
Gravitational lensing by spinning stars, approximated as homogeneous
spheres, is discussed in the weak field limit. Dragging of inertial
frames, induced by angular momentum of the deflector, breaks spherical
symmetry. I examine how the gravito-magnetic field affects image
positions, caustics and critical curves. Distortion in
microlensing-induced light curves is also considered.
\end{abstract}

\maketitle

\section{Introduction}

Gravitational lensing is one of the most extensively investigated
phenomena of gravitation and it is becoming a more and more important
tool for experimental astrophysics and cosmology.

The impressive development of technical capabilities makes it possible
to obtain observational evidences of higher-order effects in
gravitational lensing phenomena in a next future, so that, a full
treatment of lensing theory to any order of approximation is demanded.
In particular, observations of gravitational lensing phenomena could
demonstrate the inertia-influencing effect of masses in motion. In
fact, the gravito-magnetic field, predicted in 1896-1918, has not yet
a firm experimental detection.

Mass-energy currents relative to other masses generate space-time
curvature. This phenomenon, known as intrinsic gravito-magnetism, is a
new feature of general relativity and other conceivable alternative
theories of gravity and cannot be deduced by a motion on a static
background (for a detailed discussion on gravito-magnetism we refer to
Ciufolini \& Wheeler \shortcite{ci+wh95}). Gravity induced by moving
matter is related to the dragging of inertial frames and the effects
of mass currents on the propagation of light signals must be
considered. Lensing of light rays by stars with angular momentum has
been addressed by several authors with very different approaches.
Epstein
\& Shapiro \shortcite{ep+sh80} performed a calculation based on the
post-Newtonian expansion. Ib\'{a}\~{n}ez and coworkers \cite{ib+ma82,iba83}
resolved the motion equation for two spinning point-like particles,
when the spin and the mass of one of the particles were zero, by
expanding the Kerr metric in a power series of gravitational constant
$G$. Dymnikova \shortcite{dym86} studied the time delay of signal in a
gravitational field of a rotating body by integrating the null
geodesics of the Kerr metric. Glicenstein
\shortcite{gli99} applied an argument based on Fermat's principle to the
Lense-Thirring metric to study the lowest order effects of rotation of
the deflector. The listed results give a deep insight on some peculiar
aspects of spinning lenses but are very difficult to generalize.
Capozziello et al.~\shortcite{cap+al99} discussed the gravito-magnetic
correction to the deflection angle caused by a point-like, shifting
lens in the weak field regime and slow motion approximation. Asada \&
Kasai \shortcite{as+ka00} considered the light deflection angle caused
by an extended, slowly rotating lens.

In a series of papers, Sereno and co-workers developed a new approach
to address the effect of the lens spin. In Sereno
\shortcite{io02pla}, the action of the gravito-magnetic field has been
considered in the usual framework of lensing theory, i.e. {\it i)}
weak field and slow motion approximation for the lens; {\it ii)} thin
lens hypothesis \cite{sef,pet+al01}. Sereno
\& Cardone \shortcite{se+ca02} applied this formalism to consider the
gravito-magnetic contribution to the deflection angle for extended,
spherically symmetric, gravitational lenses in rigid rotation. Then,
the gravito-magnetic correction to the deflection angle has been
considered for spiral galaxy models of astrophysical interest
\cite{cap+al03}. Finally, Sereno~\shortcite{io02sub} extended the
formalism of gravitational lensing in metric theories of gravity, up
to the post-post-Newtonian order.

The simplest lens model, the point-like Schwarzschild lens, cannot
produce an intrinsic gravito-magnetic field since the local Lorentz
invariance on a static background does not account for the dragging of
inertial frames \cite{ci+wh95}. Viable theories of gravity, such as
general relativity, are classical-nonquantized theories where the
classical angular momentum of a particle goes to zero as its size goes
to zero. To treat the gravito-magnetic field, we need a further step
after the point mass as a lens model: extended lens models have to be
considered.

In this paper, I consider the lensing signatures of the homogeneous
spinning sphere. The omogeneous sphere is a usual approximation to
model several astrophysical systems, such as stars. A quite complete
theoretical analysis of the lensing properties, both inside and
outside the lens radius, is given. All the relevant lensing quantities
will be corrected for the gravito-magnetic effect. In general, the
inversion of the lens equation becomes a mathematical demanding
problem. But the gravito-magnetic effect is an higher-order correction
and interesting gravitational lens systems can be studied in some
details despite of their complexity using a perturbative approach.
This procedure is quite usual in gravitational lensing problems
\cite{ko+ko99,boz00}. By perturbing the non-rotating case, critical
curves, caustics and image positions for a rotating system will be
calculated. Our results extend previous analyses \cite{iba83,gli99}.
In Section~\ref{basic}, basic notations for spherically symmetric
lenses in rigid rotation will be introduced. In Section~\ref{homo},
the deflection angle and the deflection potential for the homogeneous
sphere will be derived. Critical curves and caustics are discussed in
Section~\ref{crit}. Section~\ref{imag} faces the inversion of the lens
mapping and present a perturbative analysis to locate the image
positions. The distortion in microlensing induced light curves is
discussed in Section~\ref{ligh}. Section~\ref{astr} gives some
estimates of the gravito-magnetic effect in the most promising
gravitational lensing system: a white dwarf as deflector. In
Section~\ref{fara}, we consider the gravitational Faraday rotation.
Section~\ref{summ} is devoted to a summary and to some final
considerations.

\section{Basics}
\label{basic}

Let us consider a class of matter distributions with a
spherically-symmetric mass density in rigid rotation. The components
of the deflection angle are \cite{se+ca02}
\begin{eqnarray}
\lab{circ1}
\hat{\alpha}_1(\xi, \theta) & =& \frac{4G}{c^2}\left\{ \frac{M(\xi)}{\xi}\cos \theta -  M(>\xi)
\frac{\omega_2}{c}\right. \\
& +& \left.
\frac{I_{\rm N}(\xi)}{\xi^2}\left( \frac{\omega_2}{c} \cos 2\theta -\frac{\omega_1}{c} \sin 2\theta \right) - \right\}; \nonumber \\
\hat{\alpha}_2(\xi, \theta)& =& \frac{4G}{c^2}\left\{ \frac{M(\xi)}{\xi}\sin
\theta +
M(>\xi)\frac{\omega_1}{c} \right. \lab{circ2} \\ & +& \left.
\frac{I_{\rm N}(\xi)}{\xi^2}
\left( \frac{\omega_1}{c} \cos 2\theta +\frac{\omega_2}{c} \sin 2\theta \right) \right\}; \nonumber
\end{eqnarray}
$c$ is the vacuum speed of light; $\xi$ and $\theta$ are polar
coordinates in the lens plane; $\omega_1$ and $\omega_2$ are the
components of the angular velocity along, respectively, the $\xi_1$-
and the $\xi_2$-axis. $M(\xi)$ is the mass of the lens within $\xi$
\cite{sef},
\beq
M(\xi)  \equiv 2 \pi \int_0^\xi \Sigma(\xi^{'}) \xi^{'} d \xi^{'},
\eeq
where $\Sigma$ is the projected surface mass density. $M(>\xi)$ is the
lens mass outside $\xi$, $M(>\xi)
\equiv M(\infty)-M(\xi)$; and
\beq
I_{\rm N}(\xi) \equiv 2 \pi \int_0^\xi \Sigma(\xi^{'}){\xi^{'}}^3 d
\xi^{'}
\eeq
is the momentum of inertia of the mass within $\xi$ about a central
axis \cite{se+ca02}. $I_{\rm N} {\times} \omega_i$ is the component of the
angular momentum along the $\xi_i$-axis.

Let us change to a dimensionless variable $\bx \equiv \bfxi /\xi_0$.
We introduce the dimensionless mass $m(x)$ within a circle of radius
$x
\equiv |\bx|$ \cite{sef},
\beq
\lab{ass8}
m(x) \equiv 2\int _0   ^{x}  k (x^{'} ) x^{'} dx ^{'} ,
\eeq
where $k$ is the convergence,
\beq
\lab{eq4}
k(\bx) \equiv \frac{\Sigma (\bx) }{\Sigma_{\rm cr}};
\eeq
$\Sigma_{\rm cr}$ is the critical surface mass density,
\beq
\lab{eq5}
\Sigma_{\rm cr} \equiv \frac{c^2}{4 \pi G} \frac{D_{\rm s}}{D_{\rm d} D_{\rm ds}}.
\eeq
being $D_{\rm ds}$ the angular diameter distance from the lens to the
source; $D_{\rm s}$ the angular diameter distance from the observer to
the source; $D_{\rm d}$ the angular diameter distance from the
observer to the deflector. We can also define a dimensionless momentum
of inertia within $x$ \cite{io02phd},
\beq
i_{\rm N}(x) \equiv 2\int _0   ^x  k (x^{'} ) x^{'3} dx ^{'} .
\eeq
With these notations, the scaled deflection angle, $\bal (\bx) \equiv
\frac{D_{\rm d} D_{\rm ds}}{\xi_0 D_{\rm s}} \hat{\bal}(\bfxi)$, can be
written as \cite{io02phd}
\begin{eqnarray}
\alpha_1(\bx )& =&
m(x) \frac{x_1}{x^2}+i_{\rm N}(x)\left[ v_2
\frac{x_1^2-x_2^2}{x^4}-v_1\frac{2x_1 x_2}{x^4}\right] \nonumber \\
& -& m(>x)v_2, \\
\alpha_2(\bx) & = &
m(x)\frac{x_2}{x^2}+i_{\rm N}(x)\left[ v_1
\frac{x_1^2-x_2^2}{x^4}+v_2\frac{2x_1 x_2}{x^4}\right]\nonumber \\
& +& m(>x)v_1.
\end{eqnarray}
where
\beq
v_1 \equiv \frac{\omega_1 \xi_0}{c},\ \ v_2 \equiv \frac{\omega_2
\xi_0}{c}
\eeq
are the circular velocities at the scale length $\xi_0$ around,
respectively, $\hat{\xi}_1$ and $\hat{\xi}_2$, in units of the speed
of light.

The scaled deflection angle is related to a dimensionless
gravitational potential $\psi$
\beq
\bfal (\bx) = \nabla \psi (\bx)
\eeq

The lens equation reads
\beq
\label{lens1}
\by = \bx - \bfal (\bx),
\eeq
where $\by$ is the position vector of the source, in units of
$\frac{D_{\rm s}}{D_{\rm d}} \xi_0$, in the source plane.

\section{The homogeneous sphere}
\label{homo}

Let us consider a homogeneous sphere of radius $R$ and volume density
$\rho_0$. The projected surface mass density is
\beq
\Sigma^{\rm HOM} (\xi)= \left\{
\begin{array}{cc}
2 \rho_0 \sqrt{R^2-\xi^2}, & {\rm if}\ \xi \leq R, \\ 0, & {\rm
elsewhere.}
\end{array} \right.
\eeq
Let us change to dimensionless variables. As a natural length scale,
we introduce the Einstein radius $R_{\rm E}$,
\beq
\xi_0= R_{\rm E} \equiv \sqrt{\frac{4 G M_{\rm TOT}}{c^2}\frac{D_{\rm d} D_{\rm ds}}{D_{\rm s}}},
\eeq
where $M_{\rm TOT}$ is the total mass of the sphere, $M_{\rm TOT}
=\frac{4}{3}\pi \rho_0 R^3$. The convergence reads
\beq
k^{\rm HOM}(x)=\frac{3}{2}\frac{1}{r^2}  \left[ 1-\left(
\frac{x}{r}\right)^2
\right]^\frac{1}{2}.
\eeq
We have
\beq
m^{\rm HOM}(x)=
\left\{
\begin{array}{cc}
1- \left[ 1-\left( \frac{x}{r}\right)^2  \right]^\frac{3}{2}, &  {\rm
if}\ x \leq r, \\ 1, & {\rm elsewhere;}
\end{array} \right.
\eeq
$r$ is the sphere radius in units of $R_{\rm E}$;
\begin{eqnarray}
i^{\rm HOM}_{\rm N}(x) & =&  \frac{2}{5}r^2  \\
& {\times} &
\left\{
\begin{array}{cc}
1- \left[ 1-\left(
\frac{x}{r}\right)^2  \right]^\frac{3}{2}
\left[ 1+\frac{3}{2}\left( \frac{x}{r}\right)^2  \right], &  {\rm if}\ x \leq r, \\
1, & {\rm elsewhere.}
\end{array} \right. \nonumber
\end{eqnarray}

Let us consider a lens rotating about the $x_2$-axis ($v_1 =0$). The
scaled deflection angle becomes \cite{io02phd}
\begin{eqnarray}
\alpha^{\rm HOM}_1(x_1,x_2) & =&  \frac{x_1}{x^2}m^{\rm HOM}(x)
 + \frac{5}{2} \frac{U}{r^2} \frac{x_1^2-x_2^2}{x^4} i^{\rm HOM}_{\rm N}(x) \nonumber \\
& - &  \frac{5}{2} \frac{U}{r^2} m^{\rm HOM}(>x), \label{angin1} \\
\alpha^{\rm HOM}_2(x_1,x_2)& =& \frac{x_2}{x^2}m^{\rm HOM}(x) + 5 \frac{U}{r^2} \frac{x_1 x_2}{x^4}i^{\rm HOM}_{\rm N}(x), \label{angin2}
\end{eqnarray}
where $U \equiv \frac{L}{c M_{\rm TOT} R_{\rm E}}$ is the ratio
between the total angular momentum of the lens, $L =\frac{2}{5}\omega
M_{\rm TOT}R^2$, and that of a particle of mass $M_{\rm TOT}$ and
velocity $c$ in a circular orbit at the Einstein radius.

Outside the lens radius ($x
> r$), the scaled deflection angle in Eqs.~(\ref{angin1},~\ref{angin2})
reduces to \cite{io02phd}
\begin{eqnarray}
\alpha^{\rm OUT}_1(x_1,x_2)& = & \frac{x_1}{x^2}+U \frac{x_1^2-x_2^2}{x^4}, \\
\alpha^{\rm OUT}_2(x_1,x_2)& = & \frac{x_2}{x^2}+2U \frac{x_1 x_2}{x^4}.
\end{eqnarray}
We will denote a deflector with the above deflection angle as the
nearly point-like, rotating lens. Its deflection angle is the same as
that of the two-point masses lens for the close binary system
\cite{ko+ko99}.

The deflection potential can be expressed as \cite{io02phd}
\beq
\psi^{\rm OUT} (x_1,x_2)=\ln x-U\frac{x_1}{x^2}.
\eeq
A point mass ($r=0$) at the origin, known as the Schwarzschild lens,
has the same lensing quantities of a homogeneous sphere outside the
radius.

\section{Critical Curves and Caustics}
\label{crit}

The Jacobian matrix of the lens mapping in Eq.~(\ref{lens1}) is,
\beq
\lab{ma5}
A(\bx )= \frac{\partial \by}{\partial \bx},\ \ \ \ A_{ij} =
\frac{\partial y_i}{\partial x_j};
\eeq
then, the magnification factor reads
\beq
\lab{ma6}
\mu ( \bx ) =\frac{1}{\det A(\bx )}.
\eeq

Points in the lens plane where the Jacobian is singular, $\det A =0$,
form closed curves, the critical curves \cite{sef,pet+al01}. They are
the locus of all images with formally infinite magnification.

The corresponding locations in the source plane are the caustics;
hence, the caustics due to a gravitational lens are the critical
values of the associated lensing map. When caustics are curves, the
smooth arcs are called folds, while cusps are the points where two
abutting fold arcs have coincident tangents with the folds arcs on
opposite sides of the double tangent \cite{sef,pet+al01}).

Let us first consider the non-rotating case, $U=0$. Multiple images
can be produced if $k(0)>1$ \cite{sef}, i.e. $r <
\sqrt{3/2}$. If $r < \sqrt{3/2}$, there are both a radial critical curve
and a tangential critical curve. The position of the radial critical
curve verifies $\frac{d}{dx}\left( \frac{m(x)}{x}\right)=1$
\cite{sef}; it is
\beq
x_{\rm r}=\frac{r}{2\sqrt{2}}\left[  (48-32r^2+r^8)^\frac{1}{2}
-r^4\right]^{1/2};
\eeq
the tangential curve is determined by $m(x)=x^2$ \cite{sef}; it is
located at
$$
x_{\rm t}=\left\{
\begin{array}{cc}
\frac{r}{\sqrt{2}}\left[ 3-r^4- (r^2-1)^\frac{3}{2}
(3+r^2)^\frac{1}{2}
\right]^\frac{1}{2}, & 1 \leq r \leq \sqrt{\frac{3}{2}}, \\
1, & r<1.
\end{array}   \right.
$$
If $r<1$, the tangential circle is superimposed to the Einstein ring.
The radial circle is inside the tangential one, $x_{\rm r} < x_{\rm
t}$.  When $r >1$, the two critical curves form inside the radius of
the sphere. When $r=\sqrt{3/2}$, they reduce to a point at the centre
of the coordinate system. For $r >\sqrt{3/2}$, no critical curve
forms.
\begin{figure}
        \resizebox{\hsize}{!}{\includegraphics{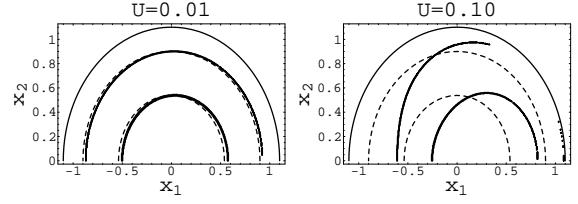}}
        \caption{The distorted tangential and radial critical curves for a
        rotating sphere with $r=1.1$. On the left panel, it is $U=0.01$; on
        the right panel, it is $U=0.1$. The thin dashed lines are the critical
        curves for a static lens; the tangential critical curve is the outer
        one. The thick lines are the distorted critical
        curves. The outer full line is the semi-circumference limiting the sphere.}
        \label{HomCriticalCurvesInside}
\end{figure}

\begin{figure}
        \resizebox{\hsize}{!}{\includegraphics{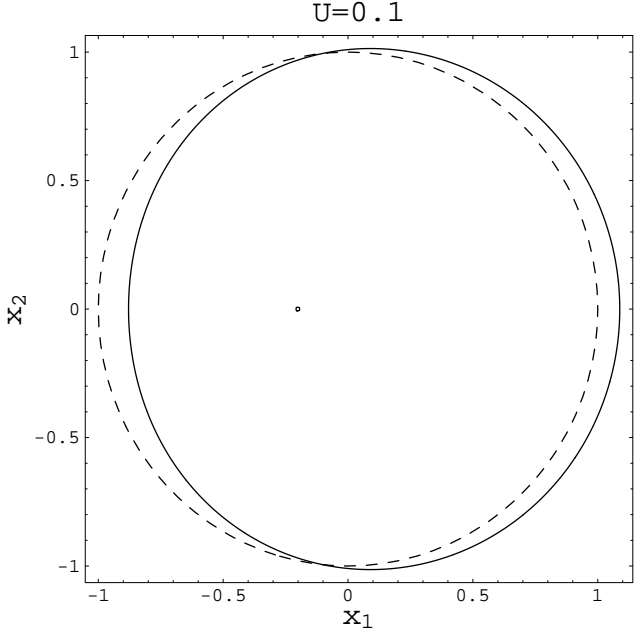}}
        \caption{The main and secondary critical curves for a nearly
        point-like rotating lens. It is $U=0.1$. The thin dashed line is the
        Einstein ring. The full curves are the critical
        curves.}
        \label{HomCriticalCurvesOutside}
\end{figure}

\begin{figure}
        \resizebox{\hsize}{!}{\includegraphics{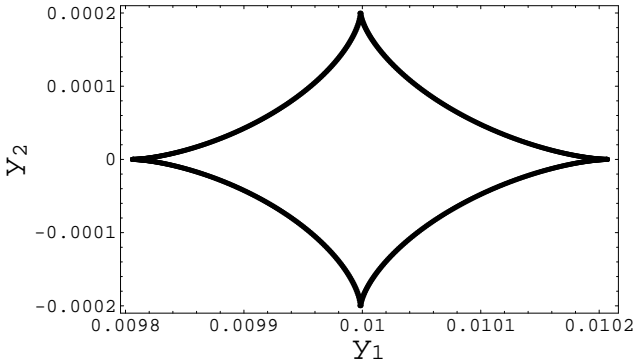}}
        \caption{The main caustic for a nearly point-like homogeneous
        sphere. It is $U=0.01$.}
        \label{HomCausticMain}
\end{figure}

The tangential critical curve is mapped on a point-like caustic at the
centre of the source plane. The radial critical curve is mapped onto a
circle. A source inside the radial caustic has three images; a source
outside the radial caustic has only one image.

The dragging of the inertial frames distorts the critical curves. In
Fig.~(\ref{HomCriticalCurvesInside}), we plot the critical curves for
a rotating sphere. Critical curves are compressed and shifted along
the $x_1$-axis. For $U>0$, the geometrical centres move to positive
values of $x_1$. Given the axial symmetry of the rotating system,
critical curves are symmetric for reflection around $\hat{x}_1$.

Let us now consider the critical curves which form outside the lens.
Outside the lens radius, the determinant of the Jacobian matrix is
\cite{io02phd}
\beq
\det A^{\rm OUT} = 1-\frac{1}{x^4}-4 U \frac{x_1}{x^6} -4 U^2 \frac{1}{x^6}.
\eeq
In the static case ($U=0$), the magnification reduces to
\beq
\mu^{\rm STAT} = \left( 1-\frac{1}{x^4}\right)^{-1}.
\eeq
For a non-rotating lens, only the tangential critical circle, in the
case $r<1$, can form outside the lens radius, when it is superimposed
to the Einstein ring. We want to discuss the distortion of the
Einstein ring induced by dragging of inertial frames. Let us perform,
first, a perturbative analysis in the hypothesis $U \ll 1$.

\begin{figure*}
        \resizebox{\hsize}{!}{\includegraphics{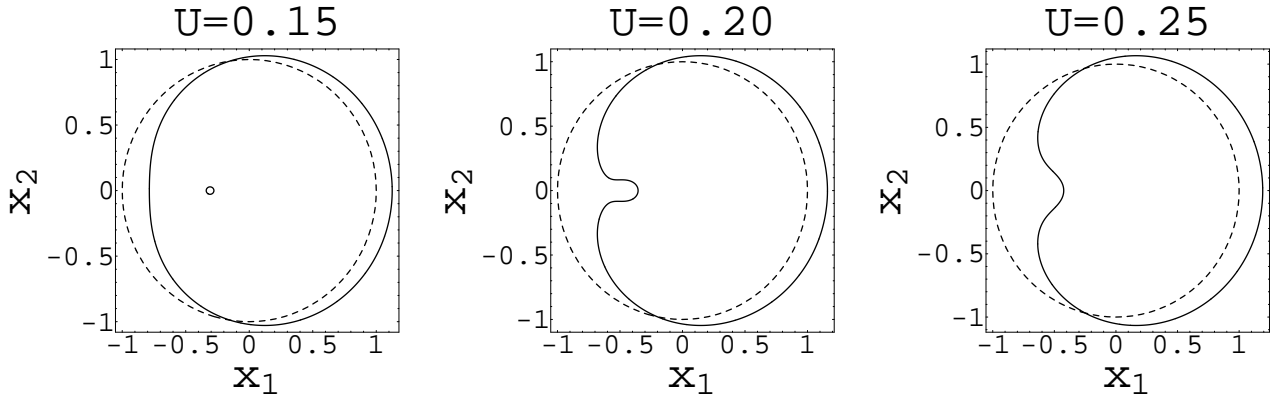}}
        \caption{The main and secondary critical curves for a nearly
        point-like rotating lens. They merge when the value of the angular momentum
        increases. On the left panel, it is $U=0.15$; on the middle panel, it is
        $U=0.20$; on the right panel, it is $U=0.25$. The thin dashed line is the
        Einstein ring. The full curves are the critical
        curves.}
        \label{HomCriticalCurvesOutsideSeries}
\end{figure*}

The equation for the main critical curve, which is a slight
modification of the Einstein circle at $x_{\rm t}=1$, is
\begin{eqnarray}
x_2(x_1) & = & \pm \left\{ -x_1^2  \right. \\
& +& \left. \left[ 54
U(U+x_1)+\sqrt{27}\sqrt{[108U(U+x_1)]^2-1} \right]^{-\frac{1}{3}}
\right. \nonumber
\\
& + & \left. \frac{1}{3}\left[ 54 U(U+x_1)+\sqrt{27}
\sqrt{[108U(U+x_1)]^2-1} \right]^\frac{1}{3} \right\} \nonumber \\
& \simeq & \pm \left\{ \sqrt{1-x_1^2} +U\frac{x_1}{\sqrt{1-x_1^2}}+U^2
\frac{1-\frac{3}{2}x_1^2}{(1-x_1^2)^{3/2}}  \right\}, \label{HomCrit}
\end{eqnarray}
where the above approximate solution in Eq.~(\ref{HomCrit}) holds for
$x_1 <1$. The main critical curve, see
Fig.~(\ref{HomCriticalCurvesOutside}), intersects the $x_1$-axis in
$x_1
\simeq
-1+U-\frac{3}{2}U^2$ and $x_1 \simeq 1+U-\frac{3}{2}U^2$. The
gravito-magnetic correction changes the width of the curve from $2$ to
$2(1-\frac{3}{2}U^2)$. The maximum height is for $x_1 \simeq U$, when
$x_2 \simeq \pm \left( 1+\frac{3}{2}U^2 \right)$; the maximum total
height changes to $\sim 2(1+\frac{3}{2}U^2)$. So, the main critical
curve is slightly compressed and its centre is shifted of $U$ along
the $x_1$-axis.

The gravito-magnetic field changes the number of critical curves.
Besides the main critical curve, a secondary critical curve forms, see
Fig.~(\ref{HomCriticalCurvesOutside}). It is centred at
$(x_1,x_2)=(-2U,0)$, and has a width $\sim {\cal{O}}(U^3)$. Since we
have considered the Jacobian for image positions outside the lens,
this secondary critical curves forms only if $2U > r$.

The main critical curve is mapped in a diamond-shaped caustic with
four cusps, see Fig.~(\ref{HomCausticMain}), which substitutes the
central point-like caustic. The main caustic is centred in
$(y_1,y_2)=( U,0)$ and its axes, parallel to the coordinate axes, are
of semi-width $\sim 2 U^2$. The secondary critical curve is mapped in
a secondary caustic, far away from the central one, centred at
$(y_1,y_2) \sim
\left(
\frac{1}{4U} -2U \sim \frac{1}{4U},0 \right)$.

When the angular momentum of the sphere increases, the two outer
critical curves merge, see
Fig.~(\ref{HomCriticalCurvesOutsideSeries}).

\section{Image positions}
\label{imag}

\begin{figure}
        \resizebox{\hsize}{!}{\includegraphics{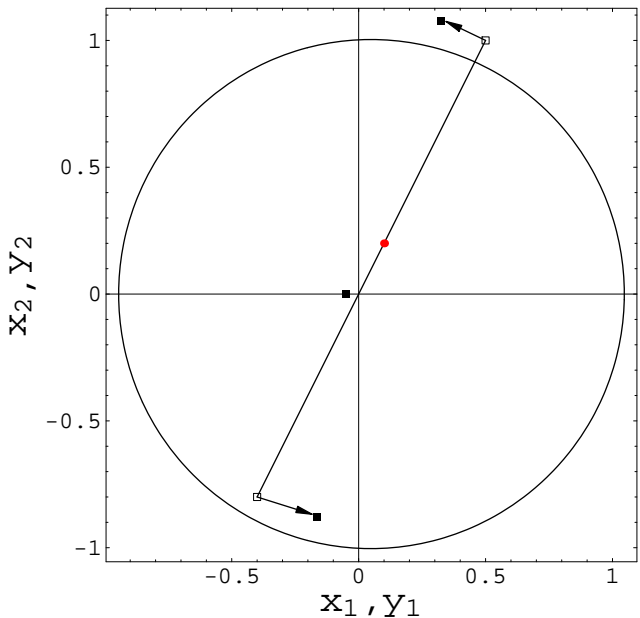}}
        \caption{Image positions for a source (grey circle) outside the caustics.
        The centre of the
        coordinate-axes, the source and the two unperturbed
        images (empty boxes) lie on a straight line. Two images (filled boxes)
        are counterclockwisely rotated,
        about the line of sight through the centre, with respect to this line;
        a third image forms near the centre. It is $U=0.05$.}
        \label{HomImagesClock}
\end{figure}

\begin{figure}
        \resizebox{\hsize}{!}{\includegraphics{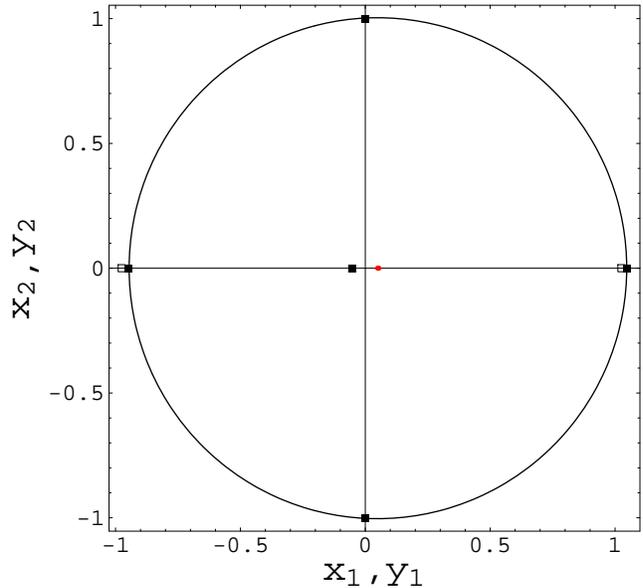}}
        \caption{Image locations for a source inside the central caustic. A
        cross pattern forms.
        The circle indicates the source position; empty boxes denote the two unperturbed
        images; the five images correspond to the filled boxes. It is $U=0.05$.}
        \label{HomImagesCross}
\end{figure}

\begin{figure}
        \resizebox{\hsize}{!}{\includegraphics{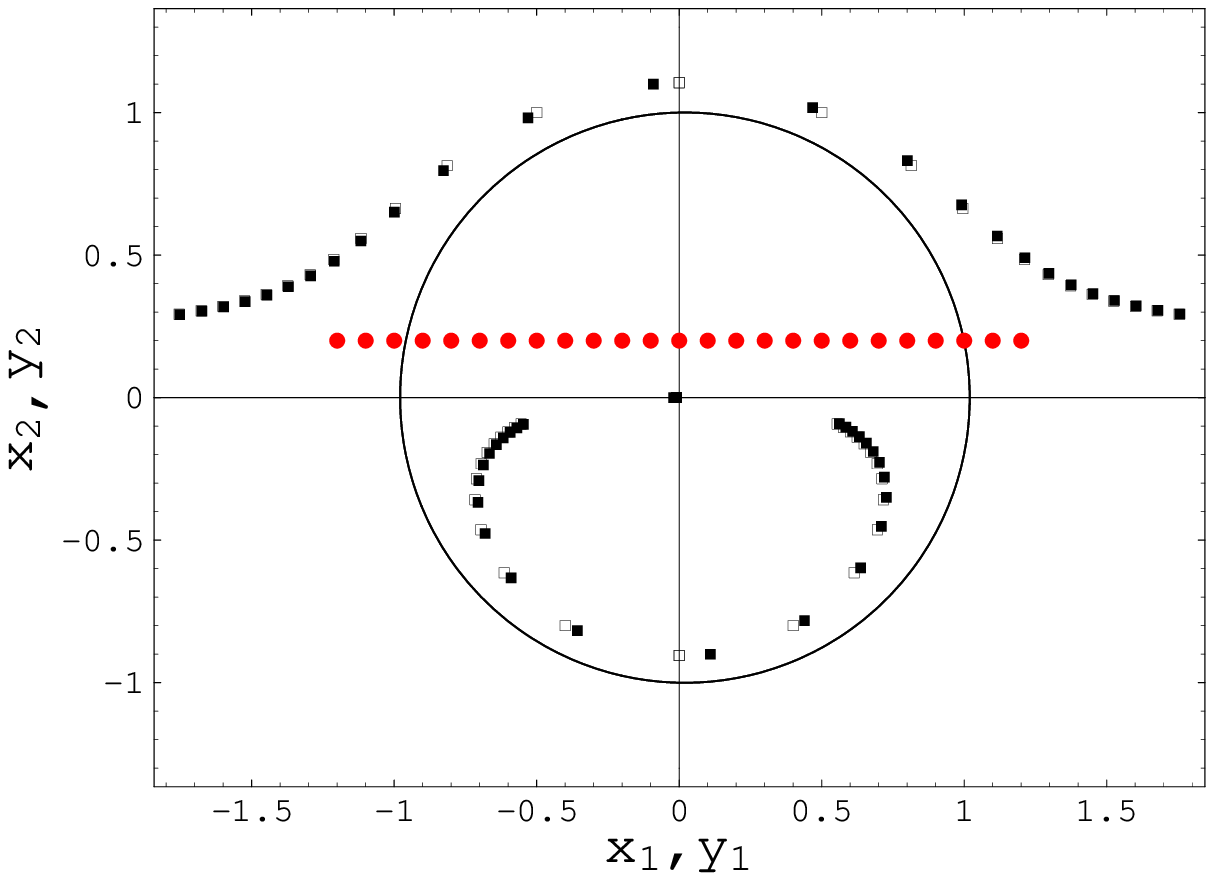}}
        \caption{A source's track, with $y_2=0.2$, and the corresponding images
        produced by a homogeneous rotating sphere. Grey circles indicates
        successive source positions. For every source
        position, two images (filled boxes) are counterclockwisely rotated, around
        the centre, with respect to the line passing through the near unperturbed
        images (empty boxes); the third image, which forms near the
        centre, is nearly stable. The main critical curve is also plotted.
        It is $U=10^{-2}$.}
        \label{HomImagesSeries}
\end{figure}

When the gravito-magnetic correction is considered, the inversion of
the lens mapping is not an easy task. However, under the condition $U
\ll 1$, we can perform a perturbative analysis. We obtain approximate
solutions to the first-order in $U$, given by
\beq
\label{pert}
\bx \simeq \bx_{(0)}+U\bx_{(1)},
\eeq
where $\bx_{(0)}$ and $\bx_{(1)}$ denote the zeroth-order solution and
the correction to the first-order. Let us consider images outside the
sphere radius.  The unperturbed images are solutions of the lens
equation for $U=0$, when the lens equation, Eq.~(\ref{lens1}), reads
\beq
\by=\bx-\frac{\bx}{x^2};
\eeq
since the potential outside the lens radius has one singularity, the
odd number images theorem is violated. Two images form at
\beq
\bx^{\pm}_{(0)}=\frac{1}{2}\left( 1 \pm \sqrt{1 +\frac{4}{y^2}}\right) \by ,
\eeq
where $y  \equiv |\by|$. The two unperturbed images lie on the
straight line going through the source position and the centre of the
coordinate system, see Fig.~(\ref{HomImagesClock}). The lensed image
$\bx_+$ lies outside the Einstein ring (on the same side of the
source), while $\bx_-$ is inside (on the side opposite the source). As
well known \cite{sef,pet+al01}, $x_+$ is a minimum and is magnified
and $x_-$ is a de-magnified saddle. As the light source moves to
infinity, $y
\rightarrow \infty$, the lensed image $x_+$ goes to infinity too and
$\mu^{\rm STAT} (x_+) \rightarrow 1$. The saddle image becomes dimmer
and dimmer, $\mu^{\rm STAT} (x_-) \rightarrow 0$, and it tends towards
the centre. If $y=0$, the source appears as an infinitely magnified
ring at the Einstein radius.

By substituting the expression in Eq.~(\ref{pert}) for the perturbed
images in the full vectorial lens equation, Eq.~(\ref{lens1}), we
obtain the first-order perturbations,
\begin{eqnarray}
x_{(1)1} & = &   \frac{x_{(0)1}^2-x_{(0)2}^2-1}{x_{(0)}^4-1},
\label{pert1}
\\ x_{(1)2} & = &
\frac{-2 x_{(0)1} x_{(0)2}}{x_{(0)}^4-1}. \label{pert2}
\end{eqnarray}
Let us consider a source on the $y_1$-axis ($y_2=0$).
Eqs.~(\ref{pert1},~\ref{pert2}) reduce to
\begin{eqnarray}
x_{(1)1} & = & \frac{1}{1+x_{(0)}^2}, \label{pert1b}
\\ x_{(1)2} & = & 0. \label{pert2b}
\end{eqnarray}
When $U>0$, i.e. when the angular momentum of the lens is positively
oriented along $\hat{x}_2$, photons with $x_1<0$ ($x_1>0$) go around
the lens in the same (opposite) sense of the deflector; photons which
impact the lens plane on the $x_1$-axis at $x_1 <0$ move closer to the
centre. This feature is common to the strong field limit of
gravitational lensing \cite{boz02}, when, again, photons in the
equatorial plane moving in the same (opposite) sense of rotation of a
black hole form closer (farther) images with respect to the
non-rotating case. If we do not limit to the equatorial plane, for
$U>0$, the images are rotated counterclockwisely around the line of
sight with respect to the non-rotating case, see
Fig.~(\ref{HomImagesClock}).

Together with these two perturbed images, a third, highly de-magnified
image, is produced near the centre, see Fig.~(\ref{HomImagesClock}).
This additional image derives from the dragging of the inertial frame;
a finite lens radius, which eliminates the central singularity, also
causes a third image. When the source is at $(-U,0)$, the third image
is superimposed to the source. The corresponding magnification factor
is $\sim {\cal{O}}(U^4)$. As can be numerically verified, for a large
range of source positions, the third image forms nearly at $(-U,0)$.
So, this image is actually outside the sphere radius only if $U
> r$, i.e $L > c M_{\rm TOT} R$; otherwise, to locate the third image,
we have to consider the lens equation with the deflection angle given
by Eqs.~(\ref{angin1},~\ref{angin2}). In usual astrophysical systems,
the condition $U > r$ is not realized, so that the third image,
forming behind the lens, is eclipsed. For a non transparent deflector,
the central image cannot be detected.

A source moving inside a caustic increases the number of images from
three to five; two additional images form. While the third central
image can be produced both by a finite size and a rotation of the
lens, these two additional images are peculiar features of the
dragging of inertial frames.

Since the gravito-magnetic effect breaks the spherical symmetry of the
system, the Einstein ring cannot form in the sky. When a source is
inside the main central caustic, four images form near the
coordinate-axes, in a cross pattern; the fifth image forms near the
centre, see Fig.~(\ref{HomImagesCross}).

In Fig.~(\ref{HomImagesSeries}), we plot the images of a source moving
in the source plane. For $U \ll 1$, two images follow the unperturbed
ones and the central image is quite stable.

\section{Light curves}
\label{ligh}

\begin{figure}
        \resizebox{\hsize}{!}{\includegraphics{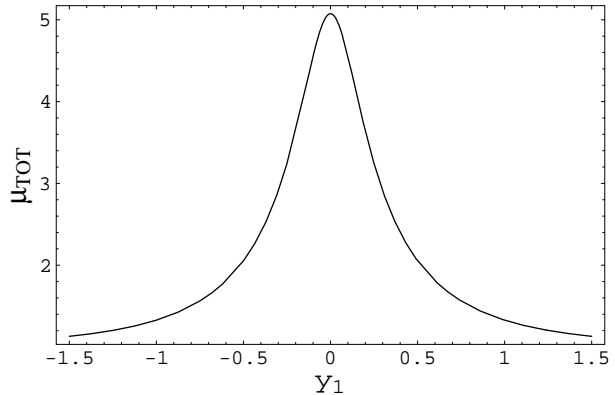}}
        \caption{The Paczy\'{n}ski curve for a source moving with $y_2=0.2$.
        The time variation is produced by the change of the source position.}
        \label{HomPac}
\end{figure}

\begin{figure}
        \resizebox{\hsize}{!}{\includegraphics{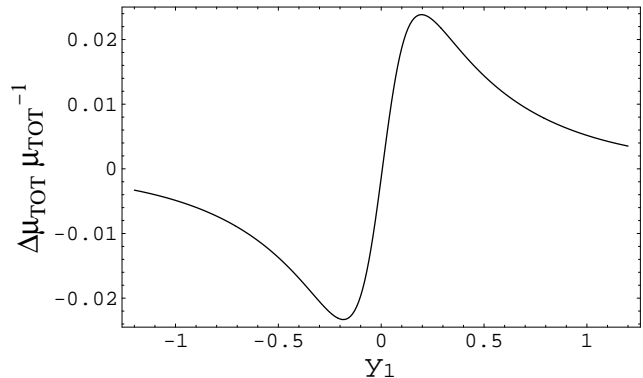}}
        \caption{The relative variation in the total light
        amplification for a point source moving with $y_2=0.2$ with respect to
        the static case. It is $U=10^{-2}$.}
        \label{HomVarMagnGRM}
\end{figure}

Let us consider distortion in microlensing-induced light curves. The
total magnification of a source is the sum of the absolute values of
the magnifications of all the images. For a non rotating sphere, if we
consider a point-like lens, the total magnification can be expressed
in terms of the source position as
\beq
\label{pac}
\mu_{\rm TOT}= \mu^{\rm STAT} (x_+) - \mu^{\rm STAT} (x_-)=\frac{y^2+2}{y \sqrt{y^2+4}}.
\eeq
When the source lies on the Einstein radius ($y=1$), the total
magnification becomes $\mu=1.34$, corresponding to a brightening by
$0.32$ magnitudes.

Unless the lens is very massive ($M > 10^6 M_\odot$ for a
cosmologically distant source), the angular separation of the two
images is too small to be resolved and is is not possible to see the
multiple images. However, a lensing event by a point mass can still be
detected if the lens and the source move relative to each other,
giving rise to lensing-induced time variability of the source. This
kind of variability, when induced by a stellar mass lens, is referred
to as microlensing. The corresponding light curve, known as
Paczy\'{n}ski curve, is described by the last term of Eq.~(\ref{pac}),
see Fig.~(\ref{HomPac}).

Let us consider how the gravito-magnetic field perturbs the
Paczy\'{n}ski curve. Numerically, the maximum relative variation, for
a source moving parallelly to the $y_1$-axis, turns out to be $\sim
\frac{1}{2} \left| \frac{U}{y_2} \right|$. In
Fig.~(\ref{HomVarMagnGRM}), the relative variation in the total
magnification is plotted. An asymmetry is induced by rotation. The
source is brightened (dimmed) when $y_1 >0$ ($y_1 <0$).

\section{Astrophysical systems}
\label{astr}

The homogeneous sphere is the standard lens model to describe lensing
by stars. As shown in Sereno \& Cardone \shortcite{se+ca02}, the most
significant case for the gravito-magnetic effect is a fast rotating
white dwarf acting as deflector. A convenient measure of the
importance of rotation is the ratio $t$ of the rotational kinetic
energy to the body's self-gravitating potential energy. For a white
dwarf, a range of values $t \sim 0.14-0.26$ can be obtained as maximum
bounds in different cases \cite{pad99}. If we take $t \sim 0.2$, it is
$L
\sim
\sqrt{0.2 G M_{\rm TOT}^3 R}$ .

The weak field limit holds when the impact parameter of the photon is
much larger than the Schwartzchild  radius, $R_{\rm Sch} =2GM/c^2$, of
the lens. Typically, for a white dwarf $M^{\rm WD}
\sim M_\odot$, $R^{\rm WD} \sim 10^{-2}R_\odot$, so that $R^{\rm WD}
\simeq 2.4 {\times} 10^3 R_{\rm Sch}^{\rm WD}$: for a photon trajectory
that just grazes the star limb, $\xi \sim R$, the weak field condition
holds. For $\xi \sim R\ (5 R)$, the rotation induces a correction of
$\sim 1.14\arcsec (0.05\arcsec)$ to the deflection angle, that is, a
correction of $\sim 0.65 \ (0.13) \%$. Astrometric accuracy of future
interferometric missions, such as SIM by NASA (scheduled for launch in
2009), should detect this correction.

Let us consider Galactic bulge-bulge lensing, when $D_{\rm s} \sim
8.5$~Kpc and $D_{\rm d} \sim 8$~Kpc; we have $U \sim 2 {\times} 10^{-7}$ and
$r \sim 2 {\times} 10^{-5} \gg U$. So, the third image is eclipsed by the lens.
For lensing of sources in the Magellanic clouds by white dwarfs in the
Halo, the situation is even worse.

The secondary critical curve and the third image form at nearly $U$,
in units of the scale length $R_{\rm E}$. The weak field limit still
holds if  $ U \gg R_{\rm s}/R_{\rm E}$, that is,  $L \gg GM^2/c $. For a
white dwarf, the angular momentum is $ \sim 3{\times}
10^{50}$~cm$^2$~g~s$^{-1}$, whereas $GM^2/c \sim
10^{49}$~cm$^2$~g~s$^{-1}$: the weak field condition is marginally
verified.

A quadrupole moment in the mass distribution of the lens also breaks
the circular symmetry. The predicted deflection due to the second
zonal harmonic, for a lens symmetric about its angular momentum and
with no angular momentum along the $\xi_1$-axis, reads
\cite{ep+sh80,io02sub},
\begin{eqnarray}
\hat{\alpha}^{J_2}_1(\mbox{\boldmath $\xi$} )
& \simeq  & 4 \frac{G M}{c^2} J_2 R^2 \left\{
\left[ 1- 4\left ( \frac{\xi_2}{\xi}\frac{L_2}{L}\right )^2 \right. \right. \nonumber \\
& -& \left. \left. \left(
\frac{L_{\rm l.o.s.}}{L} \right)^2 \right] \frac{\xi_1}{\xi} \right\}
\frac{1}{\xi^3}, \\
\hat{\alpha}^{J_2}_2(\mbox{\boldmath $\xi$} )
& \simeq  & 4 \frac{G M}{c^2} J_2 R^2 \left\{
\left[ 1- 4\left (\frac{\xi_2}{\xi}\frac{L_2}{L}\right )^2 \right. \right. \nonumber \\
& -&  \left. \left. \left(
\frac{L_{\rm l.o.s.}}{L} \right)^2 \right] \frac{\xi_2}{\xi} +
2\left(\frac{\xi_2}{\xi} \right) \left( \frac{L_2}{L}\right )^2
\right\}
\frac{1}{\xi^3},
\end{eqnarray}
where $J_2$ is the dimensionless coefficient of the second zonal
harmonic and $L_{\rm l.o.s.}$ is the component of the angular momentum
along the line of sight. The angular dependence of the
gravito-magnetic correction differs from the quadrupole effect
\cite{io02sub}. The sign of the quadrupole correction does not change
in the equatorial plane, whereas the gravito-magnetic correction
varies on opposite sides of the lens, so that it can be separated
experimentally from other terms. Furthermore, for a deflector with
angular momentum with generic direction in the space, the
gravito-magnetic effect depends only on the component of $\bf L$ in
the lens plane \cite{io02pla}, whereas $L_{\rm l.o.s.}$ enters the
quadupole correction. To evaluate the relative contribution to the
deflection angle, let us consider photon trajectories which impact the
lens plane on the $\xi_1$-axis. It is
\begin{equation}
\hat{\mbox{\boldmath $\alpha$}}^{J_2}(\mbox{\boldmath $\xi$} ) \simeq  4 \frac{G M}{c^2} J_2 R^2
\left( \frac{L_2}{L}\right)^2\frac{\hat{\xi}_1}{\xi^3}.
\end{equation}
We have to compare the above expression to the strength of the
gravito-magnetic correction, $\alpha^{\rm GRM} =\frac{4
G}{c^2}\frac{L_2}{\xi^2}$. For light rays lying in the equatorial
plane ($L_{\rm l.o.s.}=0$), it is
\begin{equation}
\left| \frac{\alpha^{J_2} }{\alpha^{\rm GRM} } \right| = J_2 \frac{c M R}{L}\frac{R}{\xi}=
J_2 \left( \frac{r}{U}\right)\left(  \frac{r}{x}\right).
\end{equation}
For the Sun, $J_2$ is order $10^{-7}$, so that the above ratio goes as
$0.25R/ \xi$. To evaluate $J_2$ for a white dwarf, let us model the
lens as a Maclaurin spheroid \cite{cha69,bi+tr87}. Then, to a ratio $t
\sim 0.2$ it corresponds an eccentricity $\sim 0.9$. $J_2^{\rm WD}$
turns out to be $\sim 0.16$ and we find a ratio of the quadrupole
correction to the gravito-magnetic effect of $\sim 24.7 R/
\xi$. In our best candidate for detecting a gravito-magnetic field,
the strength of the quadrupole correction is significant, so that the
gravito-magnetic contribution can be separated only due to its
peculiar signatures.

\section{The gravitational Faraday rotation}
\label{fara}
The plane of polarization of light rays passing close to a lens may
undergo a rotation. This is the well known gravitational Faraday
rotation \cite{ple60,ish+al88,nou99}. In a Kerr space-time, The
rotation angle, $\Omega^{\rm FAR}$, is proportional to the mass and
the line-of-sight component of the angular momentum of the deflector.
It is \cite{ish+al88,nou99}
\begin{equation}
\Omega^{\rm FAR} \sim \frac{G}{c^2}\left( \frac{R_{\rm Sch}}{\xi}\right)
\frac{L_{\rm l.o.s}}{\xi^2}.
\end{equation}

Contrary to the Faraday effect, the gravito-magnetic contribution to
the deflection angle depends only on the component of the angular
momentum in the lens plane. For equatorial trajectories of the
photons, the rotation of the polarization plane is null whereas the
gravito-magnetic contribution to the deflection angle is maximum. In
practice, however, the Faraday rotation is really small. Let us
compare its magnitude with $\alpha^{\rm GRM}$. It is
\begin{equation}
\frac{\Omega^{\rm FAR}}{\alpha^{\rm GRM}} \sim \left(
\frac{L_{\rm l.o.s.}}{L_2}\right) \left( \frac{R_{\rm Sch}}{\xi} \right)
\end{equation}
For a light ray grazing a white dwarf limb, it is $\Omega^{\rm
FAR}/\alpha^{\rm GRM} \sim 10^{-4}$.

\section{Summary and discussion}
\label{summ}

We have explored gravitational lensing by a spinning homogeneous
sphere. Some properties of the dragging of inertial frames are
independent of the specific lens model \cite{io02phd}, so that, it is
useful to not confine the analysis outside the lens. With regard to
the effect of the gravito-magnetic field, important results can be
easily generalized from the homogeneous sphere to other lens models.

Angular momentum breaks the spherical symmetry of the system leading
to interesting effects in the lensing behaviour. The effects of a
gravito-magnetic field are peculiar and can be distinguished from
other higher order corrections, such as those induced by a quadrupole
moment in the deflection potential.

The ray-trace equation for a non-rotating spherically symmetric lens
can be reduced to a one-dimensional equation but the gravito-magnetic
field breaks this symmetry. Even for very simple mass distributions,
we have to consider the full vectorial equation. However, interesting
results can be obtained either numerically or with a perturbative
approach.

A perturbative approach allows to derive simple analytical expressions
for image positions and critical curves. For a nearly point-like
spinning lens, the deflection angle has the same expression of a close
binary system. The tangential critical curve produced by a static
homogeneous sphere is distorted and shifted and a secondary critical
curve forms. The point-like central caustic changes to an extended
diamond-shaped caustic.

Photons which move in the equatorial plane in the same (opposite)
sense of the spinning lens are attracted (sent away) by the lens. For
a positively oriented angular momentum, images are counterclockwisely
shifted in the lens plane.

The total number of images is odd. With respect to the Schwarzschild
lens, a third, highly de-magnified image forms nearly behind the lens.
When the source moves into a caustic, two additional images appears.
Instead of an Einstein ring, a source inside the central caustic
produce a cross shaped pattern made of four images, nearly on the
coordinate axes, and a fifth image near the centre.

Microlensing-induced light curves are slightly distorted by the
gravito-magnetic field. The asymmetry, caused by dragging of inertial
frames, is usually negligible in realistic astrophysical lensing
systems on galactic scales. However, space telescopes performing
high-precision photometry, such as Eddington from ESA, should be able
to detect the asymmetry for microlensing events, with very small
impact parameters, induced by fast rotating white dwarf.

The weak field approximation holds whenever the distance of the light
ray to the lens is much larger than its Schwarzschild radius. This
happens for main sequence stars, early type stars and white dwarfs.
For such astrophysical situations, light rays passing close to the
Schwarzschild radius are absorbed by the lens and cannot be observed.
On the other hand, collapsed objects can probe the gravitational field
in the neighbourhood of the lens. A spinning black hole produce the
third image outside the lens, but in a region where the weak field
condition is only marginally satisfied. A full treatment of this
systems demands for an investigation in the strong field limit of
gravitational lensing. Together with the images we have discussed,
black holes produce, by strong field lensing, two infinite series of
relativistic images, formed by rays winding around the lens at
distances comparable to the gravitational radius. Bozza
\shortcite{boz02} addressed the quasi-equatorial lensing by a Kerr black hole in the strong
field limit. A comparison between the two analyses shows how the
rotation of the lens induces some features which are independent of
the lensing regime. The shift of images and the topology of critical
curves and caustics in the strong field limit are similar. In both
regimes, caustics acquire finite extension and drift away from the
optical axis. Furthermore, the asymmetry between images formed by
photons winding in the same sense of the black hole and those winding
in the opposite sense, which appear farther the lens, is preserved in
the strong filed limit. A full comparison between lensing by rotating
black holes in the weak or strong field limit cannot be performed
since the quasi-equatorial approximation misses additional images. The
general case is the objective of a future work \cite{bo+se03}.

We finally remark how a comparison among general relativity and other
viable theories of gravity can be led on the base of higher-order
effects \cite{io02phd,io02sub}. An analysis to the lowest order might
hide such differences. In this context, the analysis of the
gravito-magnetic effect deserves particular attention.

\section*{acknowledgments}
I thank Valerio Bozza for some helpful comments on the manuscript.

\end{document}